\documentclass[aps,pra,twocolumn,superscriptaddress,nofootinbib]{revtex4-1}
\usepackage{graphicx}
\usepackage{amssymb}
\usepackage{mathtools}
\usepackage{amsmath}
\usepackage{bm}
\usepackage{epstopdf}
\usepackage{xcolor}

\begin{document}

\title{Topological Floquet engineering of a 1D optical lattice via resonantly shaking with two harmonic frequencies}

\author{Jin Hyoun Kang}%\email{kangjin0119@snu.ac.kr}
\author{Yong-il Shin}\email{yishin@snu.ac.kr}
\affiliation{Center for Correlated Electron Systems, Institute for Basic Science (IBS), Seoul 08826, Korea}
\affiliation{Department of Physics and Astronomy, and Institute of Applied Physics, Seoul National University, Seoul 08826, Korea}

\begin{abstract}
We investigate the topological properties of a resonantly shaken one-dimensional optical lattice system, where the lattice position is periodically driven with two harmonic frequencies to generate one- and two-photon couplings between the two lowest orbitals. In a two-band approximation, we numerically show that degenerate edge states appear under a certain driving condition and that the corresponding topological phase is protected by the chiral symmetry of the periodically driven system. The system's micromotion is characterized with oscillating Zak phases and we find that the Zak phases are quantized only at the time when the chiral symmetry condition is explicitly satisfied.  Finally, we describe the topological charge pumping effect which arises when the driving parameters are slowly modulated around a critical point, and investigate its adiabaticity for increasing the modulation frequency.
\end{abstract}

\maketitle

\section{Introduction}

% 1. Topological insulator & Floquet TI
Floquet engineering concerns generating new properties out of a system by periodically driving it with its external or internal parameters~\cite{Goldman14, Oka19}. Under the periodic driving, the system jitters but its state can be well defined in a stroboscopic sense, thus possibly possessing effective properties which are different from those of a static system. In line with increasing interest in topological insulators and related physics~\cite{Hasan10}, Floquet engineering of band structures was actively investigated with many lattice systems~\cite{Oka09,Kitagawa10,Lindner11,Rechtsman13,Wang13sci,Zhang14,Zheng14,Jotzu14,Flaschner16}, establishing a novel direction for creating topological states. Featuring high tunability of system parameters, good isolation from environment, and low characteristic driving frequency, ultracold atoms in optical lattices present a versatile experimental platform for studying such Floquet topological insulators~\cite{Eckardt17} and also exploring their possible correlation effects~\cite{Junemann17,Barbarino19}. By generating synthetic gauge fields with Raman laser couplings~\cite{Aidelsburger13,Mancini15,Stuhl15,Han19} and controlling next-nearest-neighbor hopping via external shaking of lattice potentials~\cite{Jotzu14,Flaschner16,Kang18,Kang20}, many topological phases were realized in previous optical lattice experiments.

%,Maczewsky17,Mukherjee17

In our recent work of Ref.~\cite{Kang20}, we experimentally demonstrated a Floquet topological ladder in a shaken one-dimensional (1D) optical lattice, where the lattice position was sinusoidally modulated with a certain frequency $\omega$ to resonantly drive two-photon coupling between the two lowest orbitals~\cite{Lindner13,Weinberg15}. Because of parity conservation, the resonant shaking dominantly induces {\it site-hopping} interorbital transitions~\cite{Zheng14,Zhang14}, thus providing effective spin-orbit coupling when the two orbitals are regarded as two pseudo-spin states. The characteristic pseudo-spin winding structure of the Floquet bands was probed using interferometric measurements~\cite{Kang20}. In the work, an extension of the resonant shaking scheme was also discussed, where the lattice position is additionally modulated with frequency $2\omega$ so that {\it on-site} interorbital transitions are also induced via one-photon coupling. Because the one-photon coupling has a different topological character from the two-photon coupling, it was anticipated that the system's topological properties can be tuned by controlling the relative amplitude and phase of the two harmonic drivings and furthermore, such tunability would allow to realize topological charge pumping in the driven lattice system by slowly modulating the driving parameters~\cite{Thouless83,Wang13,Mei14,Nakajima16,Lohse16,Sun17}.

%Bukov15

In this paper, we numerically study the topological properties of the 1D optical lattice system under the two-frequency resonant driving and present a comprehensive description of the tunable Floquet topological ladder. We first examine the quasienergy spectrum and edge states of the shaken lattice system in a two-band approximation and confirm that a topologically nontrivial phase emerges under a certain driving condition, which is protected by the chiral symmetry. We characterize the system's micromotion with oscillations of the Zak phases of the Floquet bands~\cite{Zak89,KSmith93} and find that the Zak phases are quantized only at the time when the chiral symmetry requirement is explicitly satisfied. This illustrates the significance of the preferred time frame for the Floquet system~\cite{Kitagawa10}. We also perform numerical simulations of the topological charge pumping effect expected for slow modulations of the driving parameters and investigate the adiabaticity of the pumping process with increasing the modulation frequency, which provides a practical guide for the experimental realization of the charge pumping effect in the driven lattice system.

The remaining of the paper is organized as follows. In Sec.~II, we present a two-band model of the resonantly shaken 1D optical lattice system and briefly review its effective time-independent Hamiltonian. In Sec.~III, we numerically investigate the quasienergy spectrum and Floquet states of the driven system and describe its topological characteristics and micromotion. In Sec.~IV,  we present the simulation results of the topological charge pumping effect, obtained from directly calculating the time-dependent Schr{\" o}dinger equation of the slowly modulated driven system. In Sec.~V, we discuss some sensible aspects in experiment, such as adiabatic loading of the Floquet band and higher band effects, and a summary is provided in  Sec.~VI.

\section{Resonantly shaken optical lattice}

\subsection{Two-band model}

Let us consider a spinless fermionic atom in a shaken 1D optical lattice potential, $V\big(x-x_0(t)\big)$. The position $x_0(t)$ of the lattice potential can be driven by  controlling the relative phase of the laser beams forming the lattice potential. In the reference frame co-moving with the driven optical lattice, the system's Hamiltonian is given by 
\begin{equation}
{H}(t) = \frac{{p}^2}{2m_a}+V(x) - F(t){x},
\end{equation}
where $p$ is the kinetic momentum of the atom, $m_a$ is its mass, and $F(t) = -m_a\ddot{x}_0(t)$ is the inertial force arising from the driving.\footnote{Under a certain unitary transformation, the effect of shaking can be described as a vector potential in a form of $A(t){p}$. In our numerical study, we found that the inertial force description is more adequate with respect to band width renormalization.}  In the tight-binding approximation, the Hamiltonian is written as
\begin{align}
{H}(t) = & \sum_{j,\alpha}\bigg[ \epsilon_\alpha c_{j,\alpha}^\dagger c_{j,\alpha} - (t_\alpha e^{-i\theta(t)} c_{j,\alpha}^\dagger c_{j+1,\alpha}+\text{H.c.}) \nonumber \\
		   & ~~~~- F(t) a \sum_{\beta\neq\alpha} \eta_{\alpha\beta}c_{j,\alpha}^\dagger c_{j,\beta}\bigg],
\end{align}
where $c_{j,\alpha}$ ($c^\dagger_{j,\alpha}$) is the annihilation (creation) operator for the atom in the Wannier state $|j,\alpha\rangle$ on lattice site $j$ in $\alpha$ band, and $\epsilon_\alpha =\langle j,\alpha |{H}_0|j,\alpha\rangle$ and $t_\alpha =-\langle j,\alpha |{H}_0|j+1,\alpha\rangle$ with ${H}_0=\frac{{p}^2}{2m_a}+V(x)$ are the on-site energy and nearest-neighbor hopping amplitude of the $\alpha$ band, respectively. $\theta(t)= -\frac{a}{\hbar}\int_0^t dt'F(t')$ is the time-dependent Peierls phase with $a$ being the lattice spacing and $\eta_{\alpha\beta} = \frac{1}{a}\langle j,\alpha | {x}| j,\beta\rangle$ is the dimensionless dipole matrix element for on-site interorbital transition~\cite{Weinberg15}.

In this work, we are interested in a situation where the lattice position is periodically modulated with two harmonic frequencies, $\omega$ and $2\omega$, i.e., $x_0(t) = x_\omega \cos (\omega t)+x_{2\omega} \cos (2\omega t+\varphi)$, which gives
\begin{eqnarray}
F(t) &=& F_{\omega} \cos(\omega t)+F_{2\omega} \cos(2\omega t+\varphi) \\
\theta(t) &=& -\theta_\omega \sin(\omega t)-\theta_{2\omega} \sin(2\omega t+\varphi) 
\end{eqnarray}
with $F_{\nu}=m_a\nu^2 x_\nu $ and  $\theta_{\nu} = a m_a \nu x_\nu /\hbar$. The modulation frequency $2\omega$ is set to be close to the energy difference between the two lowest, $s$ and $p$ orbitals, so strong band hybridization arises from their resonant couplings via one-`photon' process for the $2\omega$ driving as well as two-`photon' process for the $\omega$ driving [Fig.~1(a)]. We assume that the two bands are energetically well separated from higher bands and the effects of their coupling to the higher bands are negligible in the system. This condition can be achieved by tailoring the optical lattice potential, for example, with a double well structure (see Sec.V). 

Taking a two-band approximation in Eq.~(2), we obtain the Bloch Hamiltonian of the driven system as
\begin{align}
{\mathcal{H}}(q,t) =& \Big( \bar{\epsilon}-2{t}_+\cos[q-\theta(t)]\Big)\mathbb{I}-F(t)a\eta_{sp}  \sigma_x \nonumber\\
			&+\Big( \epsilon-2t_-\cos[q-\theta(t)]\Big)\sigma_z
\end{align}
for ${\Psi}_q = (c_{q,p}, c_{q,s})^\text{T}$, where $c_{q,s(p)}$ is the annihilation operator for the state with quasimomentum $q$ in the $s(p)$ band. Here $2\bar{\epsilon} = \epsilon_p +\epsilon_s$,  $2\epsilon=\epsilon_p-\epsilon_s$, $2{t}_+ = t_p+t_s$, $2t_- = t_p-t_s$, $\mathbb{I}$ is the identity matrix, $\bm{\sigma}=\{\sigma_x,\sigma_y,\sigma_z\}$ are the Pauli matrices, and $q$ is expressed in units of $1/a$.

%%%%%%% Figure 1 %%%%%%%
\begin{figure}[t]
\includegraphics[width=7.0cm]{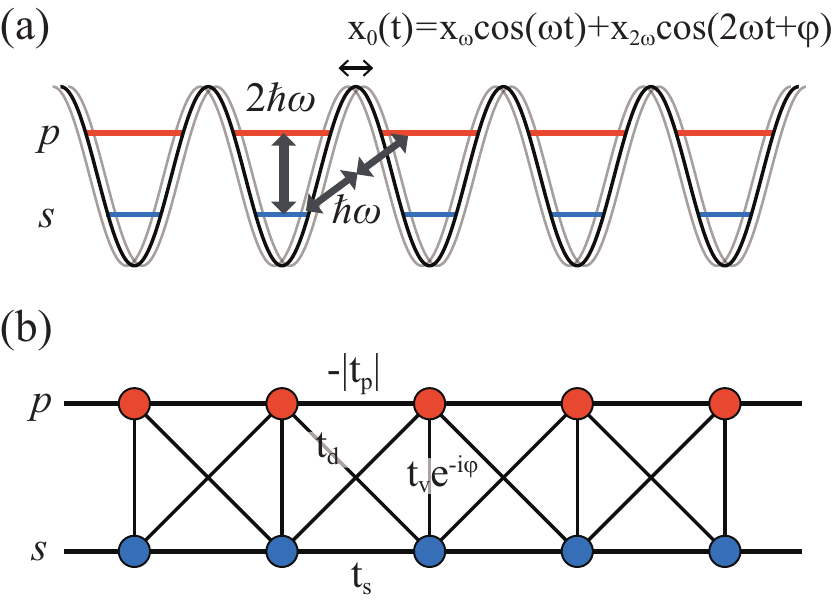}
\centering
\caption{Resonantly shaken 1D optical lattice. (a) Schematic of the shaken lattice system. The lattice position is modulated with two harmonic frequencies, $2\omega$ and $\omega$, to generate one-photon and two-photon resonant couplings between the $s$ and $p$ orbitals, respectively. (b) Effective ladder model of the driven lattice system. The two orbitals comprise the two legs of the ladder and the vertical and diagonal interleg links are formed with hopping amplitudes, $t_v e^{-i\varphi}$ and $t_d$ by the one- and two-photon resonant couplings, respectively.   $t_{\alpha \in \{s,p\}}$ denotes the intraleg hopping amplitude of $\alpha$ band.}
\end{figure}
%%%%%%%%%%%%%%%%%%%

\subsection{Effective Hamiltonian}

Before proceeding to a numerical investigation of Eq.~(5), we derive the effective time-independent Hamiltonian ${\mathcal{H}}_\text{eff}$ of the resonantly driven lattice system. To handle the resonant interband coupling suitably, we first apply a unitary transformation of ${U}_R(t) = \text{exp}(-i\omega t \sigma_z)$ to the Bloch Hamiltonian ${\mathcal{H}}(q,t)$~\cite{Goldman15}, yielding 
\begin{align}
{\mathcal{H}}'(q,t) =& \Big( \bar{\epsilon}-2{t}_+\cos[q -\theta(t)]\Big)\mathbb{I}-F(t) a \eta_{sp} e^{2i\omega t \sigma_z}\sigma_x \nonumber\\
			&-\Big(\Delta +2t_-\cos[q -\theta(t)]\Big)\sigma_z
\end{align}
with $\Delta=\hbar\omega-\epsilon$. In this rotating frame, the driving frequency $\omega$ is larger than any other relevant energy scale of the system and thus, the effective Hamiltonian can be perturbatively obtained using the high-frequency expansion method~\cite{Rahav03, Eckardt15}. When the Fourier series expansion of ${\mathcal{H}}'(q,t)$ is given as ${\mathcal{H}}'(q,t) = \sum_m {\mathcal{H}}_m(q) e^{im\omega t}$, the second-order approximation of  ${\mathcal{H}}_\text{eff}$ is given by
\begin{align}
{\mathcal{H}}_\text{eff}(q) &= {\mathcal{H}}_0 +\sum_{m>0} \frac{[{\mathcal{H}}_m,{\mathcal{H}}_{-m}]}{m\hbar\omega} \nonumber \\
&= \bar{E}(q)\mathbb{I} - \bm{h}(q)\cdot \bm{\sigma}
\end{align}
and we obtain
\begin{eqnarray}
\bar{E}(q) &=& \bar{\epsilon}-2{t}'_+ \cos(q) \\
\bm{h}(q) &=& \big(\Delta'+2t'_- \cos(q) \big)\hat{\bm{z}} + 2t_d\sin(q) \hat{\bm{y}}  \nonumber\\
&&  +~ t_v \big(\cos(\varphi) \hat{\bm{x}}+ \sin(\varphi) \hat{\bm{y}}\big) 
\end{eqnarray}
with ${t}'_+={t}_+\mathcal{J}_0(\theta_\omega)$, $\Delta'=\Delta- \hbar\omega \eta_{sp}^2 (\frac{\theta_\omega^2}{3}+\frac{\theta_{2\omega}^2}{4})$, $t'_- = t_-\mathcal{J}_0(\theta_\omega)$, $t_d = \theta_\omega\eta_{sp}t_-\mathcal{J}_1(\theta_\omega)$, $t_v =\hbar\omega \theta_{2\omega}\eta_{sp}$, and $\mathcal{J}_n$ being the $n$-th order Bessel function of the first kind. The details of the derivation is provided in Appendix A.

The effective interband coupling in the driven system is revealed by the transverse component of $\bm{h}(q)$. As indicated by $t_d\propto \theta_\omega \eta_{sp} t_-$, the second term, $2t_d \sin (q) \hat{\bm{y}}$, is derived from the two-photon interorbital transition induced by the $\omega$ driving. Because the on-site two-photon transition between the $s$ and $p$ orbitals is forbidden due to parity conservation, the dominant two-photon process involves site hopping, so the corresponding transverse field has sinusoidal $q$ dependence. The third term in Eq.~(9) with $t_v\propto \theta_{2\omega} \eta_{sp}$ is a uniform transverse field, which results from the on-site one-photon interorbital transition by the $2\omega$ driving. We note that the direction of the uniform field is determined by the relative phase $\varphi$ of the two harmonic drivings.

The effective system has chiral symmetry for $\varphi = \pm\frac{\pi}{2}$,  as $\sigma_x {\mathcal{H}}_\text{eff}(q)\sigma_x = -{\mathcal{H}}_\text{eff}(q)$.\footnote{The term $\bar{E}(q)\mathbb{I}$ in Eq.~(7) is ignored, which can be cancelled out under a proper gauge transformation without affecting the topological properties of the system.} Geometrically, it means that the $\bm{h}$ vector is confined in the $yz$ plane, and two topologically distinctive phases may exist for the system, which are characterized by the winding of $\bm{h}$ around the origin. When the following conditions are satisfied,
\begin{equation}
|\Delta'/t'_-|<2 ~~\text{and}~~|t_v/t_d|<2,
\end{equation}
the trajectory of $\bm{h}(q)$ as $q$ traverses the Brillouin zone (BZ) encircles the origin, indicating nontrivial band topology, and otherwise, $\bm{h}(q)$ shows no winding, corresponding to a topologically trivial phase. It is important to note that by breaking the chiral symmetry with $\varphi \neq \pm\frac{\pi}{2}$, the two topologically distinct phases can be continuously connected to each other in the parameter space spanned by $\{t_d,t_v,\varphi\}$. This is the key feature of the driven lattice system that allows the topological charge pumping effect~\cite{Sun17}.

The effective system with ${\mathcal{H}}_\text{eff}$ can be viewed as a cross-linked two-leg ladder system, as sketched in Fig.~1(b). The two legs represent the $s$ and $p$ orbitals, and the direct and diagonal interleg links denote the interorbital couplings induced by the $2\omega$ and $\omega$ resonant drivings, respectively. In past studies, a two-leg cross-linked ladder system under a magnetic field, referred to as Creutz ladder~\cite{Creutz99}, was discussed as a minimal model for 1D topological insulators~\cite{Junemann17,Hugel14}. In our ladder system, $t_s t_p<0$ is equivalent to having a $\pi$ gauge flux per ladder plaquette and the relative driving phase $\varphi$, which appears as the complex phase of the direct link, can be regarded as determining the subplaquette flux distribution. Thus, the two-band system under the two-frequency resonant driving effectively realizes a generalized Creutz ladder with tunable interleg links.

\section{Floquet state analysis}

To elucidate its topological properties and real-time dynamics, we perform a Floquet state analysis on the driven two-band system~\cite{Eckardt17}. Floquet states $|\psi_n(t)\rangle$ are eigenstates of the time-evolution operator over one driving period $T=\frac{2\pi}{\omega}$, i.e., 
\begin{equation}
{U}(t+T,t)|\psi_n(t)\rangle =e^{-i\varepsilon_n T/\hbar}|\psi_n(t)\rangle,
\end{equation}
where ${U}(t+T,t) = \mathcal{T}\text{exp}[-\frac{i}{\hbar}\int_{t}^{t+T} {H}(t')dt']$ with $\mathcal{T}$ being the time-ordering operator. $\varepsilon_n \in [-\frac{\hbar\omega}{2},\frac{\hbar\omega}{2})$ is referred to as quasienergy for the Floquet state and its spectrum is independent of the choice of time $t$. The Floquet states provide a proper basis set for describing the system's dynamics in a stroboscopic manner, i.e., when the Floquet system is prepared in a state $|\psi\rangle=\sum_n c_n |\psi_n (0)\rangle$ at $t=0$, its state at $t=mT$ with $m$ being integer is given by $|\psi(t)\rangle=\sum_n c_n e^{-i \varepsilon_n m T} |\psi_n(0)\rangle$.  

We numerically calculate the time-evolution operator as ${U}(t+T,t) = \prod_{n=0}^{N-1} {U}(t_{n+1}, t_n)$  with ${U}(t_{n+1}, t_n) = \text{exp}[-\frac{i}{\hbar}{H}(t_n)\Delta t]$, $t_n = \frac{nT}{N}+t$, and $\Delta t = \frac{T}{N}$ for large $N$, and determine the quasienergy spectrum and Floquet states, $\{\varepsilon_n,|\psi_n(t)\rangle\}$, by directly diagonalizing ${U}(t+T,t)$ for various driving conditions as well as time $t$. In our calculations, we consider a lattice system of $L=150$ sites and the lattice parameters are set to be $\epsilon_s = 0$, $\epsilon_p=4.4 E_r$, $t_s = 0.03 E_r$, $t_p=-0.31 E_r$, $\hbar\omega =\epsilon=2.2 E_r$, and $x_\omega=0.11a$ with $E_r = \frac{\hbar^2\pi^2}{2m_a a^2}$ as in the previous experiment~\cite{Kang20}.

\subsection{Edge states} 

%%%%%%% Figure 2 %%%%%%%
\begin{figure}[t]
\includegraphics[width=8.4cm]{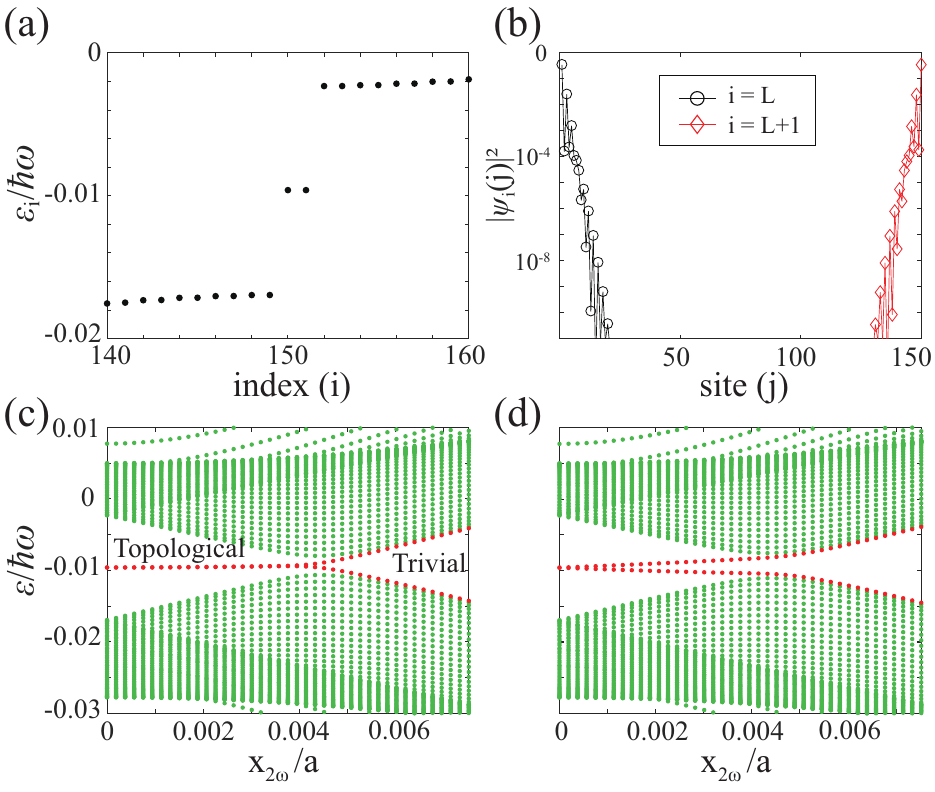}
\centering
\caption{Floquet states of the resonantly shaken 1D lattice system. (a) Quasienergy spectrum $\{\varepsilon_i\}$ for $x_\omega = 0.11 a$ and $x_{2\omega}=0$. The spectrum was obtained for a lattice system of $L = 150$ sites with open boundary. (b) Probability density profiles of the degenerate in-gap states in (a). Evolution of the quasienergy bands as functions of $x_{2\omega}$ for (c) $\varphi =0.5\pi$ and (d) $\varphi =0.45\pi$. $a$ denotes the lattice spacing. In (c), a topological phase transition occurs as $x_{2\omega}$ increases over a certain critical value $x_{2\omega,c}$.}
\end{figure}
%%%%%%%%%%%%%%%%%%%

We first investigate a case where the system is driven only with frequency $\omega$. For this single frequency driving, we have $t_d\neq 0$ and $t_v=0$, and the effective Hamiltonian $\mathcal{H}_\text{eff}$ predicts emergence of a topologically nontrivial phase which is characterized by the winding of $\bm{h}$. We observe that two degenerate in-gap states exist in the quasienergy spectrum [Fig.~2(a)] and also that their wavefunctions are exponentially localized at both edges of the system [Fig.~2(b)]. According to the bulk-edge correspondence, these degenerate edge states are a hallmark of 1D topological phase, confirming the topological effect of the two-photon resonant coupling in the driven lattice system. 

We further examine how the system evolves as the second harmonic, $2\omega$ driving is additionally applied to it. Figure 2(c) displays the evolution of the quasienergy spectrum as a function of the driving amplitude, $x_{2\omega}$, for $\varphi=\frac{\pi}{2}$. As $x_{2\omega}$ increases, the band gap is reduced to close and when $x_{2\omega}$ exceeds a certain critical value $x_{2\omega,c}$, it becomes reopened. Meanwhile, the in-gap states maintain their degeneracy for $x_{2\omega}<x_{2\omega,c}$. These observations clearly show that a topological phase transition occurs at $x_{2\omega}=x_{2\omega,c}$ with increasing $x_{2\omega}$ in the driven system. The critical driving amplitude is measured to be $x_{2\omega,c}=0.0044 a$ for $x_{\omega}=0.11 a$, which is found to be consistent with the critical point of $t_v/t_d=2$ from the effective system with ${\mathcal{H}}_\text{eff}$.\footnote{$\Delta'/t'_-\approx 0$ for our driving condition.}

In Fig.~2(d), we display the evolution of the quasienergy spectrum for different $\varphi =0.45\pi$. Different from the case of $\varphi=\frac{\pi}{2}$, it is observed that the edge state degeneracy is immediately lifted for $x_{2\omega}\neq0$. We checked that it is the case for any value of $\varphi\neq \pm \frac{\pi}{2}$. In the effective Hamiltonian description, it is explained as a result of breaking the chiral symmetry which protects the topological phase. Thus, we infer that the topological phase of the resonantly driven lattice system is protected by the same chiral symmetry. The chiral symmetry of the Floquet system is expressed as 
\begin{equation}
\sigma_x {\mathcal{H}}(q,t+t_0) \sigma_x = -{\mathcal{H}}(q,-t+t_0)
\end{equation}
with a proper choice of time frame $t_0$~\cite{Kitagawa10}. For the given forms of ${F}(t)$ and $\theta(t)$ in Eqs.~(3) and (4), respectively, we find that the symmetry condition is satisfied only with $\varphi = \pm\frac{\pi}{2}$ (mod $2\pi$) at $t_0 = \pm\frac{T}{4}$ (mod $T$).\footnote{The identity matrix term of ${\mathcal{H}}(q,t)$ in Eq.~(5) is ignored with the same reason explained in the discussion of ${\mathcal{H}}_\text{eff}$.} The significance of the preferred time frame will be clear in the following subsection regarding the micromotion of the driven lattice system.

\subsection{Micromotion}

While the quasienergy spectrum of a Floquet system is time-independent, the system's Floquet state changes with time, which corresponds to the jittering micromotion of the periodically driven system. Here we consider an insulating state of the driven lattice system, where one of the Floquet bands are fully occupied by atoms and investigate its micromotion in terms of current oscillations. The current of atoms in the filled Floquet band is given by~\cite{KSmith93,Resta00,Xiao10} 
\begin{align}
j_{\pm}(t)= & \frac{i}{2\pi} \int_\text{BZ} dq~ \partial_{t}\langle \psi_q^{\pm}(t)|\partial_q|\psi_q^{\pm}(t)\rangle, 
\end{align}
where $|\psi_q^{\pm}(t)\rangle$ is the Floquet state with quasimomentum $q$ at time $t$, and it is determined from the time evolution operator ${U}(t+T,t;q) = \mathcal{T}\text{exp}[-\frac{i}{\hbar}\int_{t}^{t+T}{\mathcal{H}}(q,t')dt']$.  The indices $\pm$ denote the upper and lower Floquet bands, respectively. With the Zak phase defined as $\gamma_{\pm} =i\int_\text{BZ}dq\langle \psi_q^{\pm}|\partial_q|\psi_q^{\pm}\rangle$~\cite{Zak89},  the current is related as $j_{\pm}(t)= \frac{1}{2\pi} \partial_{t}\gamma_{\pm} (t)$ and thus, the micromotion of the insulating state is characterized by the temporal evolution of the Zak phase of the Floquet band.

%%%%%%% Figure 3 %%%%%%%
\begin{figure}[t]
\includegraphics[width=8.4cm]{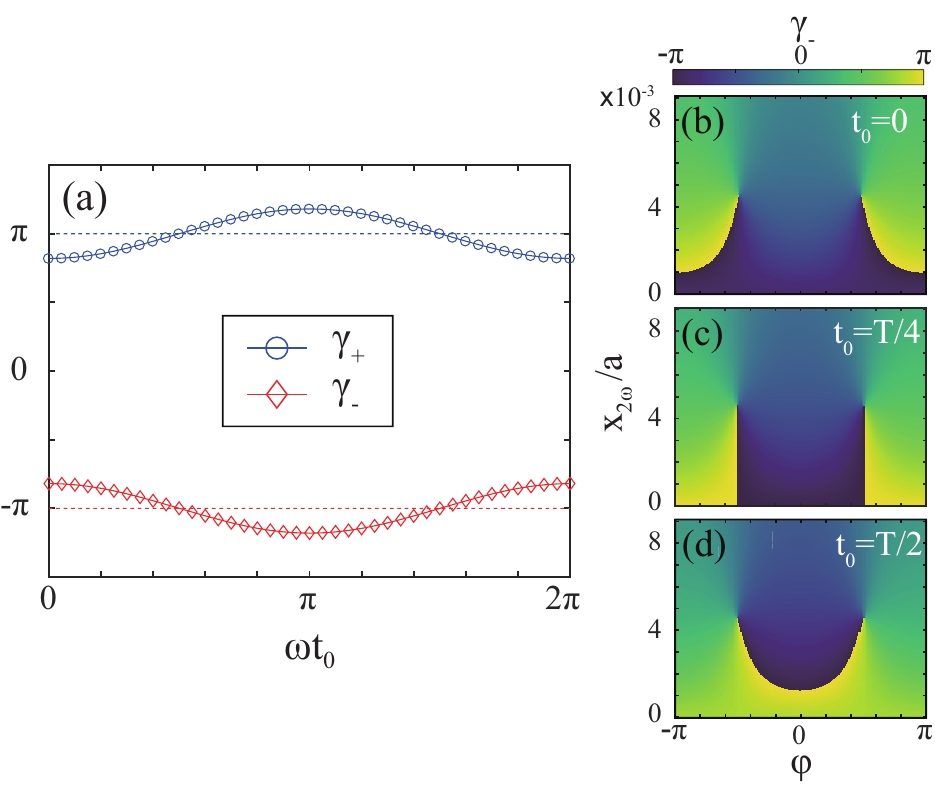}
\caption{Oscillations of the Zak phases, $\gamma_\pm$, of the Floquet bands. (a) Temporal evolution of $\gamma_\pm$ over one driving period, $0<t<T=\frac{2\pi}{\omega}$, for $\{x_\omega,x_{2\omega},\varphi\}=\{0.11 a,0,\frac{\pi}{2}\}$. $\pm$ denotes the upper and lower Floquet bands, respectively. Zak phase $\gamma_-$ in the plane $\{x_{2\omega},\varphi\}$ at (b) $t_0=0$, (c) $\frac{T}{4}$, and (d) $\frac{T}{2}$. The singular points are always located at $(x_{2\omega,c},\pm\frac{\pi}{2})$.}
\end{figure}
%%%%%%%%%%%%%%%%%%%

In Fig.~3(a), we display $\gamma_\pm(t)$ for $x_{2\omega}=0$, where the system is in a topologically nontrivial phase having a pair of degenerate edge states [Fig.~1(a)]. The Zak phases exhibit non-zero values and oscillate around $\gamma_\pm=\pm \pi$ with the same period of $T$. The sum of the two Zak phases is always zero, meaning that the currents in the two Floquet bands cancel each other. We observe that $\gamma_\pm$ have $\pm \pi$ values, i.e., are quantized only at $t=\frac{T}{4}$ and $\frac{3T}{4}$. Remarkably, these time points indeed correspond to the preferred time frame $t_0=\pm \frac{T}{4}$ (mod $T$) for the chiral symmetry in Eq.~(12). The concurrence of the Zak phase quantization and the symmetry condition fulfillment shows that the effective Hamiltonian ${\mathcal{H}}_\text{eff}(q)$ obtained from the high-frequency expansion method is closely associated with the Floquet Hamiltonian ${\mathcal{H}}_F (q,t)$ at $t=t_0$, which is defined as ${\mathcal{H}}_F(q,t)=\frac{i}{T}\log [{U}(t+T,t;q)]$.

In Figs.~3(b)--3(d), we plot $\gamma_-$ in the plane of $x_{2\omega}$ and $\varphi$ at $t = 0$, $\frac{T}{4}$, and $\frac{T}{2}$, respectively. Singular points appear at $\{x_{2\omega,c},\varphi=\pm\frac{\pi}{2}\}$ regardless of $t$ and the Zak phase exhibits $2\pi$ winding around each of the critical points. We see the branch cut at $\gamma_-=\pm\pi$ move over time, indicating the system's micromotion under the period driving. At $t=\frac{T}{4}$, the Zak phase distribution in the driving parameter space is found to be identical to that for ${\mathcal{H}}_\text{eff}$.

% corroborating the suggestion of ${\mathcal{H}}_\text{eff}(q)={\mathcal{H}}_F(q,t_0)$.

\section{Topological charge pumping}

%\subsection{Slowly modulated driving}

When the driving parameters $\{x_{2\omega}, \varphi\}$ slowly vary on the time scale of the driving period, the system can {\it adiabatically} follow the change of the driving condition, i.e., the system's long-time dynamics is governed by the time-varying effective Hamiltonian ${\mathcal{H}}_\text{eff}(q;t)={\mathcal{H}}_\text{eff}(q; \{x_{2\omega}(t), \varphi(t)\})$~\cite{Weinberg17,Novicenko17}. This adiabatic following implies that when the driven lattice system is prepared in a Floquet insulating state, it can be adiabatically manipulated into another insulating state, particularly, with different topological properties. 

Based on this adiabatic following, topological charge pumping is realized in the driven lattice system when the driving parameters are slowly modulated around a critical point. As a general example, let us consider a specific pumping protocol with $x_{2\omega}(t) =  x_{2\omega,c}\big(1+ 0.43\cos(\frac{2\pi t}{T_p})\big)$ and  $\varphi(t) = \frac{\pi}{2}\big(1+\frac{1}{2}\sin(\frac{2\pi t}{T_p})\big)$  for $x_\omega =0.11 a$, where $T_p$ is the pumping cycle time [Fig.~4(a)]. In the $x_{2\omega}$--$\varphi$ plane, the state point of the periodic driving revolves around the critical point at $(x_{2\omega,c}, \frac{\pi}{2})$ in clockwise. When the driven system prepared in an insulating phase, for example, in the lower Floquet band, it will experience a Zak phase change of $2\pi$ every cycle~[Fig.~3(c)], which results in unity charge transportation because $\int_0^{T_p} j_-(t') dt' = \frac{1}{2\pi} [\gamma_-(T_p)-\gamma_-(0)]=1$. In other words, when the system returns to its initial state after one pumping cycle, all the atoms are shifted by one lattice site. This charge pumping process is a topological one in that it occurs only when the driving parameter trajectory encloses the critical point in the 2D parameter space~\cite{Thouless83,Wang13,Mei14,Nakajima16,Lohse16,Sun17}. In Fig.~4(b), the evolution of the quasi-energy spectrum of the driven system during the pumping cycle is displayed, which clearly shows propagating edge modes in the bulk gap, an analogue of Chern insulator in the $q$--$t$ plane~\cite{Resta00,Xiao10}.  

%%%%%%% Figure 4 %%%%%%%
\begin{figure}[t]
\includegraphics[width=8.4cm]{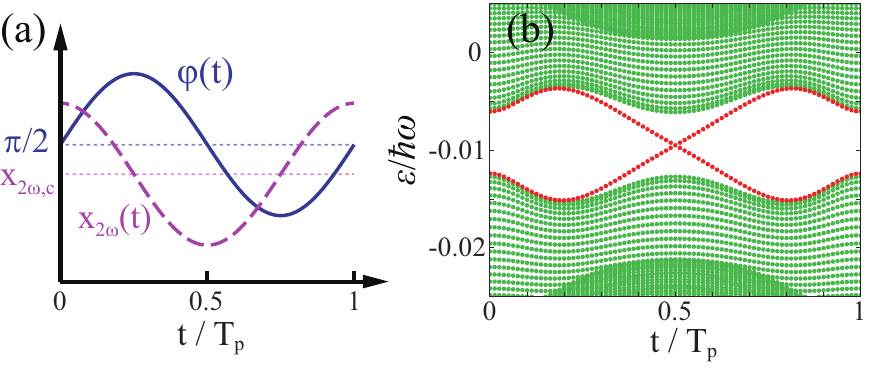}
\centering
\caption{
Charge pumping protocol for the resonantly driven lattice system. (a) The driving parameters are periodically modulated as $x_{2\omega}(t) = x_{2\omega,c}[1+0,43\cos(2\pi t/T_p)]$ and $\varphi(t) = \frac{\pi}{2}[1+ \frac{1}{2}\sin(2\pi t/T_p)]$. In the $x_{2\omega}$--$\varphi$ plane, the state point of the periodic driving revolves around the critical point at $(x_{2\omega,c}, \frac{\pi}{2})$ in clockwise. (b) Evolution of the quasienergy bands of the driven lattice system over the modulation period. The edge modes propagate between the bulk bands during the pumping cycle.}
\end{figure}
%%%%%%%%%%%%%%%%%%%

We test the topological charge pumping effect by directly calculating the time-dependent Schr{\" o}dinger equation of the driven system under the cyclic modulations of the driving parameters.  The initial state is set to be the insulating state of the lower Floquet band, $|\psi(q,0)\rangle=|\psi_q^-(0)\rangle$ and the time evolution of $|\psi(q,t)\rangle$ is calculated from $i\partial_t |\psi(q,t)\rangle={\mathcal{H}}(q,t) |\psi(q,t)\rangle$ with the time-varying driving parameters, $x_{2\omega}(t)$ and $\varphi(t)$. The current is obtained as $j (t)= \frac{1}{2\pi}\int_\text{BZ} \langle \psi(q,t)|{v}(q,t)|\psi(q,t)\rangle$ with velocity operator ${v}(q,t) = \partial {\mathcal{H}}(q,t) /\partial (\hbar q)$~\cite{Xiao10,Sun17} and its time integration gives the pumped charge amount as $C(t) = \int_0^t dt' j(t')$.

%%%%%%% Figure 5 %%%%%%%
\begin{figure}[t]
\includegraphics[width=8cm]{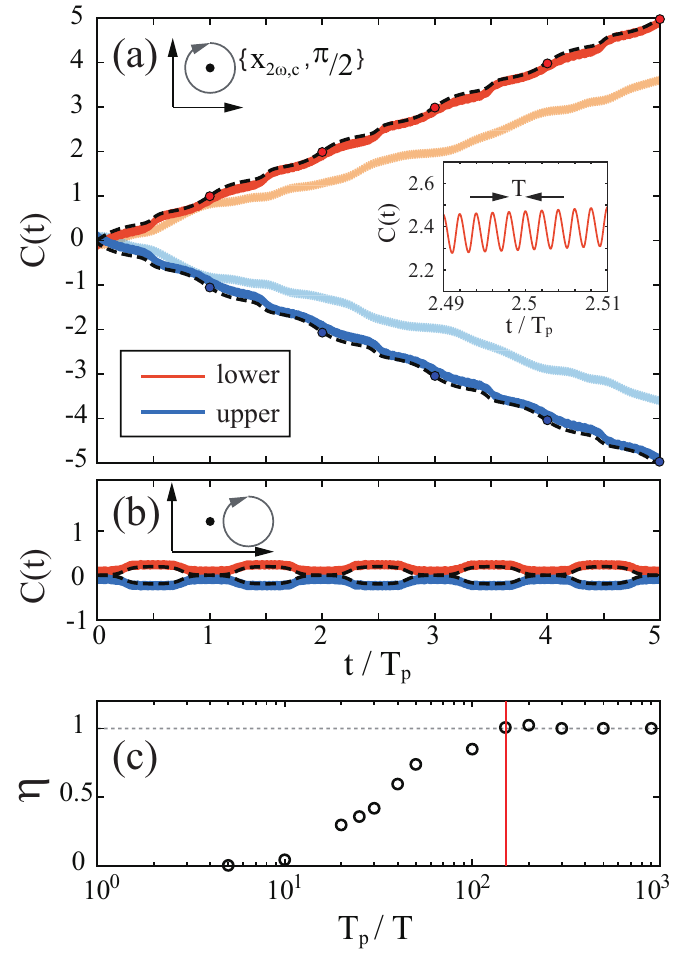}
\centering
\caption{Numerical simulation of the topological charge pumping effect. (a) Pumped charge amount, $C(t)$, as a function of the pumping time $t$ for the pumping protocol described in Fig.~4(a), with $T_p=900T$. The red and blue solid lines indicate the results for the lower and upper Floquet bands, respectively, and the black dashed lines denote the predictions from the time-varying effective Hamiltonian ${\mathcal{H}}_\text{eff}(q;t)$. The inset shows the short-time behavior of $C(t)$, revealing the micromotion of the driven system. The faint lines are the results obtained with short $T_p=50T$. (b) Numerical results for a modified pumping sequence with $\varphi(t) = \frac{\pi}{4}\sin(2\pi t/T_p)$ ($T_p=500T$), where the pumping trajectory in the $x_{2\omega}$--$\varphi$ plane does not enclose the critical point. (c) Charge pumping rate per cycle, $\eta$, as a function of $T_p$, obtained by averaging over 5 pumping cycles. The red vertical line indicates the characteristic time scale of the driven lattice system, determined by its Floquet band gap energy [Fig.~4(b)].}
\end{figure}
%%%%%%%%%%%%%%%%%%%

In Fig.~5(a), we display the pumped charge $C(t)$ as a function of time for $T_p = 900 T$. It is observed that $C(t)$ increases by unity in every pumping cycle, i.e., $C(m T_p)=m$, as expected. The inset shows the short-time evolution of $C(t)$, where it reveals fast oscillations with the time period of $T$, which indicates the system's micromotion under the fast-periodic driving. The red dashed line denotes the prediction from the time-varying effective Hamiltonian ${\mathcal{H}}_\text{eff}(q;t)$ and it is found to well describe the time evolution of the envelope of $C(t)$. This demonstrates that the driven lattice system adiabatically follows the slow modulations of the driving parameters with such long $T_p\gg T$. In addition, we check that when the system is prepared in the upper Floquet band, the charge pumping direction is reversed for $\gamma_+=-\gamma_-$ [Fig.~5(a)] and also that when the driving parameter trajectory does not enclose the critical point, for example, with $\varphi(t) = \frac{\pi}{4}\sin(\frac{2\pi t}{T_p})$, there is no net charge transport after one cycle [Fig.~5(b)].

%\subsection{Adiabatic pumping}

The adiabaticity of the charge pumping process is investigated by performing the simulations with reducing the modulation period $T_p$. The pumped charge after 5 pumping cycles is measured and the pumping efficiency is determined as $\eta=\frac{C(5T_p)}{5}$ for various $T_p$~[Fig.~5(c)]. We observe that the pumping efficiency is reduced from unity as $T_p$ decreases below about $200 T$ and it eventually vanishes when $T_p< 10 T$. When the change of the driving parameters is not slow enough, atoms cannot remain in the original Floquet band and they would make Landau-Zener-type interband transitions during the pumping process. For our given pumping protocol, the band gap energy of the driven lattice system is estimated to be $\Delta_{g}\approx 0.01 \hbar \omega$ [Fig.~4(b)] and it explains the observed time scale $T_p\approx 100T$ for the adiabaticity of the charge pumping process. 

\section{Discussions}

\subsection{Floquet band loading}

In a typical optical lattice experiment, ultracold atoms are initially prepared in the lowest band of a static lattice potential. Because the bare $s$ band of the static lattice potential is topologically trivial, one may be concerned that experimentally realizing the Floquet topological insulating state would be practically inefficient because it is inevitable to close the band gap to transfer the initial state into the topologically nontrivial state~\cite{Dauphin17}. However, as illustrated in the topological charge pumping process, in the two-frequency driving scheme, the protecting symmetry can be broken by varying $\varphi$, so the gap can be maintained to be open in the band transfer from trivial to topological by properly designing the path in terms of the driving parameters including the driving frequency $\omega$. In the effective Hamiltonian description, the band gap energy is given by $\Delta_g=\min\big[ 2|\bm{h}(q)| \big] $ with
\begin{align}
|\bm{h}(q)|=&  \big[ (\Delta'+ 2t'_- \cos q)^2 + 4 t_d^2 \sin^2q \nonumber \\ &+4 t_d t_v \sin \varphi \sin q + t_v^2 ]^{\frac{1}{2}}.
\end{align}
The adiabatic transfer from the initial $s$-band insulating state to a Floquet topological state, for example, with $\Delta'=0$, $t_d\neq0$, and $t_v=0$ can be achieved by the following steps: (1) first setting the driving frequency as $|\Delta'|>2|t'_-|$ with $t_d=t_v=0$, (2) increasing $t_v$ to a finite magnitude, (3) decreasing $|\Delta'|$ to zero, (4) ramping up $t_d$ to the target value with $\varphi=0$, and (5)  finally, reducing $t_v$ to zero. In Fig.~6, we illustrate the evolution of the band structure for the loading sequence. Note that in the initial setting of $\omega$, $\Delta'>-\frac{\epsilon}{2}$ to prevent two-photon resonant transition by $2\omega$ driving.

 %%%%%%% Figure_add %%%%%%%
\begin{figure}[t]
\includegraphics[width=8.5cm]{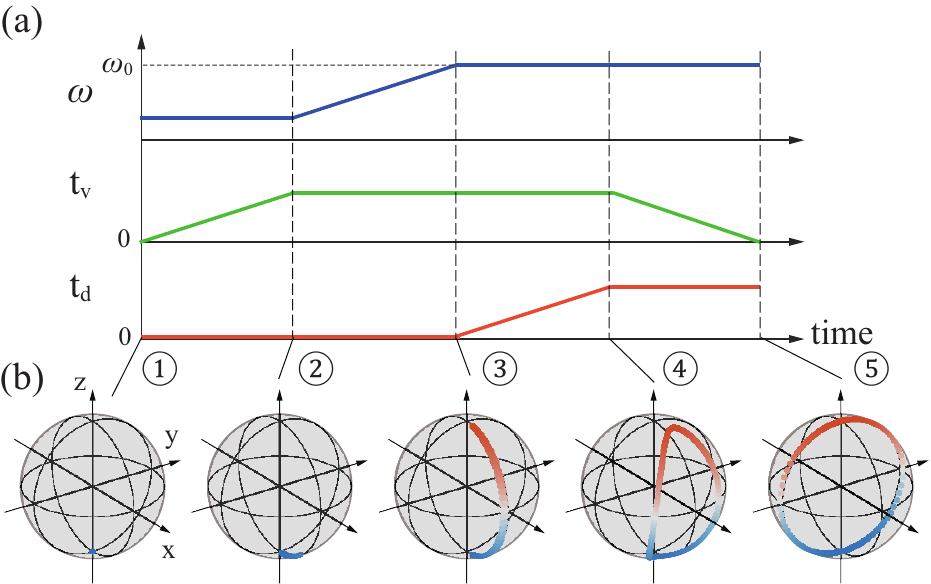}
\caption{Adiabatic formation of the topologically non-trivial Floquet bands. (a) Time sequence of the experimental control of the lattice driving parameters. $\omega_0$ denotes the resonance frequency such that $\Delta'=0$. (b) Corresponding evolution of the pseudo-spin trajectory of the ground band on the Bloch sphere. The north and south poles of the sphere represent the $p$ and $s$ orbital states, respectively. The initial, bare $s$ band is continuously transformed into a topologically non-trivial band, whose pseudo-spin trajectory makes a great circle on the Bloch sphere.}
\end{figure}
%%%%%%%%%%%%%%%%%%%

\subsection{Higher band effect}

The two-band approximation was adopted in this work, requiring energetic isolation of the two lowest bands in the lattice system. However, in a generic optical lattice potential of a sinusoidal form, the energy differences between orbitals are comparable to each other, so it can be difficult to fully validate the two-band approximation. The participation of the higher bands in the system's dynamics might interfere the formation of a topological phase. In the long time limit, it would cause heating and atom loss in the driven lattice system, which was an important subject in recent studies~\cite{Dalessio14,Weidinger17,Reitter17,Boulier19,Singh19}.

In our previous experiment, we performed Ramsey-type interferometric measurements using two separate pulses of lattice shaking to probe the $\bm{h}$-field distribution in the momentum space~\cite{Kang20}. The measurement results clearly revealed the winding of $\bm{h}(q)$ for the two-photon resonant coupling but it was found that the measured interferometric signals were not fully explained by an effective two-band description. Because the experiment employed a sinusoidal optical lattice potential, it was pointed out that the higher band effect might be a source of the deviation.

 %%%%%%% Figure 6 %%%%%%%
\begin{figure}[t]
\includegraphics[width=6.5cm]{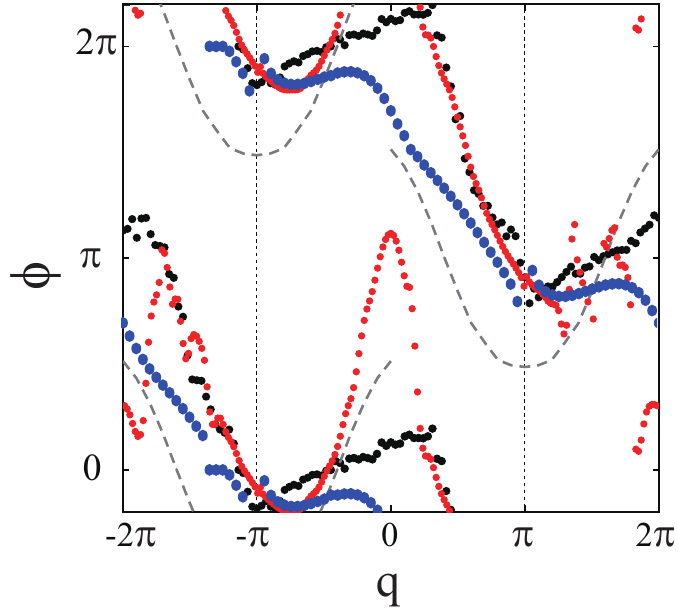}
\caption{Effectiveness of the two-band descriptions of a resonant shaking experiment using a sinusoidal optical lattice potential. The black dots indicate the experimental result of Ref.~\cite{Kang20}, where the oscillation phase of Ramsey fringes was measured as a function of quasimomentum $q$ with a resonantly shaken 1D optical lattice system. Here, $|q|<\pi$ and $\pi<|q|<2\pi$ correspond to the $s$ and $p$ bands, respectively. Three different models were used to numerically simulate the experiment result: (grey dashed line)  the effective two-band model with $\mathcal{H}_\text{eff}(q,t)$, (blue dots) the time-dependent two-band model with $\mathcal{H}(q,t)$, including the micromotion dynamics, and (red dots) direct calculation of $H(t)$ in Eq.~(1), taking into account the higher bands.}
\end{figure}
%%%%%%%%%%%%%%%%%%%

To investigate it, we numerically simulate the measurement experiments with three different models using (A) the effective two-band Hamiltonian $\mathcal{H}_\text{eff}(q,t)$, (B) the time-dependent two-band Hamiltonian $\mathcal{H}(q,t)$, and (C) a multi-band description based on the plane wave analysis of $H(t)$ (see Appendix B), respectively. Figure 7 shows the simulation results in comparison with the experimental data. It is clear that the model B, which includes the micromotion of the driven system, gives an improved description of the experimental data, without predicting such a phase jump observed at $q=0$ in the model A~\cite{Goldman15}. Noticeably, the numerical result from the model C shows good quantitative agreement with the experimental data in the ranges of $0.2\pi<q<\pi$ and $-1.6\pi<q<-\pi$. Although its discrepancy near $q=0$ is still large even for the inhomogeneous trapping potential which is not taken into account in the model, it seems to suggest that the coupling to the higher bands was not negligible in the experiment.

%%%%%%% Figure 7 %%%%%%%
\begin{figure}[t]
\includegraphics[width=6.0cm]{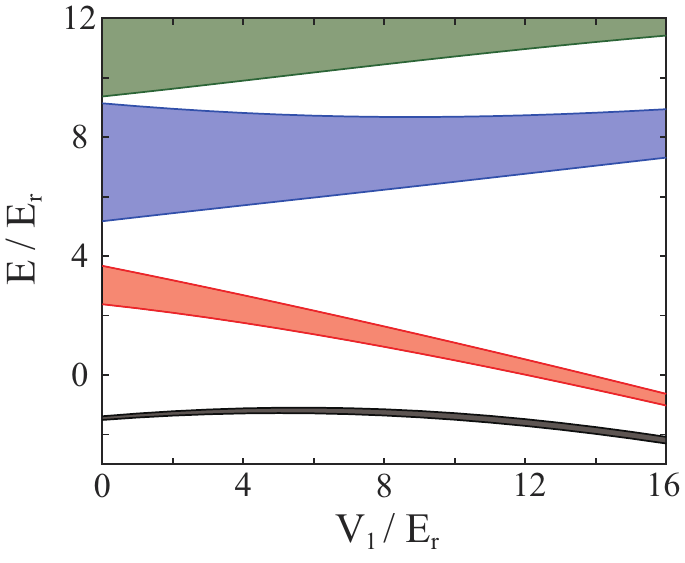}
\caption{Evolution of the energy bands of the superlattice potential $V(x)$ in Eq.~(15) as a function of $V_1$ for $V_0 = 8E_r$. $E_r = \frac{\hbar^2\pi^2}{2m_a a^2}$ is the recoil energy of the lattice potential.}
\end{figure}
%%%%%%%%%%%%%%%%%%%

One of the ways to mitigate the higher band effect is tailoring the optical lattice potential into a superlattice form like 
\begin{align}
V(x) =& \frac{V_0}{2} \cos \bigg(\frac{2\pi}{a} x \bigg)+\frac{V_1}{2} \cos \bigg(\frac{4\pi}{a} x \bigg).
\end{align}
The unit cell has a double-well structure so that the $s$ and $p$ orbitals are in pair separated from the other orbitals. In Fig.~8, we display the evolution of the energy band structure of the superlattice potential as a function of $V_1$. With increasing the depth of the short lattice potential, $V_1$, the $p$ band is pulled toward the $s$ band while the $d$ and $f$ bands are pushed further away from them. As the strength of $n$-photon resonant interband coupling is approximately proportional to $\theta_\omega^n$~\cite{Weinberg15}, it is clear that the higher band effect will be suppressed by using such superlattice potentials. For example, when $V_0=V_1=8E_r$ and $\hbar\omega = \epsilon$, the $p$--$d$ interband coupling strength is estimated to be less than $1\%$ of that of the $s$--$p$ coupling up to $\theta_\omega<1.4$, where the $p$--$d$ coupling is generated via $n=4,5,6$ transitions.

\section{Summary}

We investigated the topological properties of the 1D optical lattice under two-frequency resonant driving and confirmed the emergence of a symmetry-protected topological phase under a certain driving condition. We showed that the oscillating Zak phases of the Floquet bands, representing the system's micromotion, are quantized at the preferred time frame when the chiral symmetry condition is explicitly satisfied. This revealed that the effective Hamiltonian ${\mathcal{H}}_\text{eff}(q)$ obtained from the high-frequency expansion method represents the Floquet Hamiltonian ${\mathcal{H}}_F (q,t)$ at the preferred time frame. We also demonstrated the topological charge pumping effect under slow modulations of the driving parameters and investigated the adiabaticity of the pumping process in the driven lattice system. This work corroborated the characteristics of the tunable Floquet topological ladder synthesized in a resonantly shaking 1D optical lattice potential and provided a practical guide for its experimental realization.

\begin{acknowledgments}
We thank Jeong Ho Han for discussion. This work was supported by the Institute for Basic Science in Korea (IBS-R009-D1) and the National Research Foundation of Korea (NRF-2018R1A2B3003373, NRF-2019M3E4A1080400).
\end{acknowledgments}

\section*{Appendix}
\subsection{High-frequency expansion}
\setcounter{equation}{0}
\numberwithin{equation}{section}
\renewcommand{\theequation}{A\arabic{equation}}

From Eq.~(6), the Bloch Hamiltonian of the driven lattice system is given by
\begin{align}
{\mathcal{H}}'(q,t) =& \Big( \bar{\epsilon}-2{t}_+\cos[q -\theta(t)]\Big)\mathbb{I}-\hbar\omega \tilde{F}(t)  \eta_{sp} e^{2i\omega t \sigma_z}\sigma_x \nonumber\\
			&-\Big(\Delta +2t_-\cos[q -\theta(t)]\Big)\sigma_z  \\
{\text{with}}~~\theta(t) &= -\theta_\omega \sin(\omega t)-\theta_{2\omega} \sin(2\omega t+\varphi) \nonumber \\
\tilde{F}(t) &= \theta_\omega \cos(\omega t)+2 \theta_{2\omega} \cos(2\omega t+\varphi). \nonumber
\end{align}
In the Fourier series of ${\mathcal{H}}'(q,t) = \sum_m {\mathcal{H}}_m(q) e^{im\omega t}$, the coefficients $\mathcal{H}_m$ are  given by
\begin{align}
{\mathcal{H}}_0  =& \Big( \bar{\epsilon}\mathbb{I}-\Delta \sigma_z\Big) -2\mathcal{J}_0(\theta_\omega)\cos(q)\Big( {t}_+\mathbb{I}+t_-\sigma_z\Big) \nonumber \\
			     & - \hbar\omega \theta_{2\omega}\eta_{sp} e^{-i \varphi \sigma_z} \sigma_x \nonumber\\
{\mathcal{H}}_{1,3} =& -2i\mathcal{J}_{1,3}(\theta_\omega)\sin(q)\Big( {t}_+\mathbb{I}+t_-\sigma_z\Big)-\hbar\omega\frac{\theta_\omega}{2}\eta_{sp}\sigma^+ \nonumber\\
{\mathcal{H}_{4}} =&  -2\mathcal{J}_4(\theta_\omega)\cos(q)\Big( {t}_+\mathbb{I}+t_-\sigma_z\Big) - \hbar\omega \theta_{2\omega}\eta_{sp} e^{i \varphi}  \sigma^+\nonumber\\
{\mathcal{H}}_{m=\text{o}} &= -2i\mathcal{J}_m(\theta_\omega)\sin(q)\Big( {t}_+\mathbb{I}+t_-\sigma_z\Big) \nonumber\\
{\mathcal{H}}_{m=\text{e}} &= -2\mathcal{J}_m(\theta_\omega)\cos(q)\Big( {t}_+\mathbb{I}+t_-\sigma_z\Big)
\end{align}
with $\sigma^+ = \sigma_x + i \sigma_y$ and $\mathcal{J}_m$ being the $m$-th order Bessel function of the first kind. Here, in the Fourier expansion of $\cos [q-\theta(t)]$, we neglect the $2\omega$ driving term in $\theta(t)$, assuming $\theta_\omega \gg \theta_{2\omega}$. o and e indicate the odd and even numbers, respectively, which are not 0, 1, 3, nor 4.

Using the high-frequency expansion method~\cite{Rahav03, Goldman15,Eckardt15}, the time-independent effective Hamiltonian $\mathcal{H}_\text{eff}(q)$ can be perturbatively obtained in a form of ${\mathcal{H}}_\text{eff} = \sum_{k=0}^\infty {\mathcal{H}}^{(k)} (\frac{1}{\hbar \omega})^k $ and the coefficient of the leading terms are given with 
\begin{align}
{\mathcal{H}}^{(0)} =& {\mathcal{H}}_0, \\
{\mathcal{H}}^{(1)} =& \sum_{m\neq 0} \frac{{\mathcal{H}}_m {\mathcal{H}}_{-m}}{m}, \nonumber \\
{\mathcal{H}}^{(2)} = & \sum_{m\neq 0}\bigg( \frac{[{\mathcal{H}}_{-m},[{\mathcal{H}}_0,{\mathcal{H}}_m]]}{2m^2} \nonumber \\
			&~~~~~+\sum_{m'\neq 0,m} \frac{[{\mathcal{H}}_{-m'},[{\mathcal{H}}_{m'-m},{\mathcal{H}}_m]]}{3mm'}\bigg). \nonumber
\end{align}
The effective Hamiltonian truncated to the first order terms is given as
\begin{align}
\mathcal{H}_\text{eff}(q) =  \mathcal{H}_0&- 2\theta_\omega\eta_{sp}t_-\mathcal{J}_1(\theta_\omega)\sin(q)\sigma_y\nonumber\\
				&+\hbar\omega \eta_{sp}^2\Big[\frac{\theta_\omega^2}{3}+\frac{\theta_{2\omega}^2}{4}\Big]\sigma_z,
\end{align}
where the higher order Bessel functions are neglected because $\mathcal{J}_1(\theta_\omega)\gg \mathcal{J}_{3}(\theta_\omega)$ for small $\theta_\omega$. In the vector form of $\mathcal{H}_\text{eff}(q) = \bar{E}(q)\mathbb{I} - \bm{h}(q) \cdot \bm{\sigma}$,
\begin{align}
\bar{E} =& \bar{\epsilon}-2{t}_+\mathcal{J}_0(\theta_\omega)\cos(q) \\
h_x =& t_v \cos(\varphi)  \nonumber \\
h_y=& t_v\sin(\varphi)+2t_d\sin(q) \nonumber\\
h_z =& \Delta'+2t_- \mathcal{J}_0 (\theta_\omega)\cos(q) \nonumber
\end{align}
with $t_v = \hbar\omega \theta_{2\omega}\eta_{sp}$, $t_d = \theta_\omega \eta_{sp}t_-\mathcal{J}_1(\theta_\omega)$, and $\Delta'=\Delta-\hbar\omega \eta_{sp}^2 [\frac{\theta_\omega^2}{3}+\frac{\theta_{2\omega}^2}{4}]$.

\subsection{Plane wave analysis}
\setcounter{equation}{0}
\numberwithin{equation}{section}
\renewcommand{\theequation}{B\arabic{equation}}

Under a unitary transformation of 
\begin{align}
{U}(t) = \text{exp} \bigg[ \frac{i}{\hbar}\int_0^t d\tau F(\tau) {x}\bigg],
\end{align}
the Hamiltonian in Eq.~(2) is transformed to 
\begin{align}
{H}'(t) = \frac{1}{2m_a}[{p} + \frac{\hbar}{a}Q(t)]^2 +\frac{V_0}{2}\cos\Big(\frac{2\pi}{a}x\Big)
\end{align}
with $Q(t) = \frac{a}{\hbar} \int_0^t d\tau F(\tau)$. Since the transformed Hamiltonian maintains translational symmetry, its eigenstates are Bloch states, $|u_q^n\rangle$, which can be expanded by plane waves as
\begin{align}
|u_q^n\rangle = \sum_{P=-\infty}^\infty c_{q,P}^n |u_{q,P}\rangle, 
\end{align}
where $n\in\{0, 1, 2, \cdots\}$ is the band index, $ \langle x|u_{q,P}\rangle = L^{-\frac{1}{2}} e^{i\frac{q}{a}x} e^{i\frac{2\pi P}{a} x}$, and $P\in \mathbb{Z}$. In the plane wave basis, the matrix representation of $H'(t)$  is given by~\cite{Weinberg15}
\begin{align}
\langle u_{q',P'} | {H}'(t)|u_{q,P}\rangle 
= \delta_{q'q}\bigg[ &\frac{E_r}{\pi^2}\big(q+2\pi P +Q(t)\big)^2 \delta_{P'P} \nonumber \\ &+\frac{V_0}{4}(\delta_{P',P+1}+\delta_{P',P-1})\bigg].
\end{align}

This provides a frame for the multi-band description introduced in Sec.~V B, referred to as model C, to investigate the higher band effect in the Ramsey-type interferometric measurements of Ref.~\cite{Kang20}. In the experiment, a Fermi gas of atoms was initially prepared in the $s$ band of a 1D optical lattice potential and two pulses of the resonant shaking were applied to the system with a variable intermission time of $T_e$, where the first shaking induced two-photon coupling and the second one generated one-photon coupling. Ramsey fringe signals were observed in the band populations and the oscillation phase $\phi(q)$ was measured in a momentum-resolved manner. In the numerical investigation, we determined $Q(t;T_e)$ from the experimental sequence and directly solved the Schr{\" o}dinger equation of the system for the initial $s$-band state of $|u_q^0\rangle$. The final state $|\psi_q(T_e)\rangle$ for the hold time $T_e$ was calculated and its $n$-th band population was determined as $n(q;T_e) = |\langle u_q^n|\psi_q (T_e)\rangle|^2$ as a function of $T_e$. The oscillation phase $\phi(q)$ of the interference signal was extracted from a sinusoidal function fit to $n(q;T_e)$ as in the experiment. In our calculations, Bloch states are expanded with 51 plane waves, i.e., $|P|\leq 25$ and it was checked that the result was not significantly changed with increasing the basis size.

\end{document}